**Construction of an Inducible TBCE Overexpression System to Probe Tubulin Cofactor–Mediated Microtubule Homeostasis**

Faiza Hasib Khan

Bio 460 mentor: Dr. Steven James

2nd reader: Dr. Ryan Kerney

3rd reader: Dr. Peter Fong

## Abstract

We are studying a new-to-science regulator of microtubule homeostasis, *wdA*, in the fungus *Aspergillus nidulans*. wdA is a 7-WD-repeat protein that forms a β-propeller structure and likely acts to scaffold a multi-protein complex. Mutations in *wdA* confer cold-sensitive lethal mitotic catastrophe, suggesting that it is important for normal MT-based processes. Microtubule function depends on a tightly regulated supply of soluble αβ-tubulin heterodimers, which are produced and maintained through the action of several tubulin-specific cofactors. Among these, TBCA and TBCE play distinct but closely connected roles in tubulin folding, assembly, and dissociation. TBCA binds β-tubulin and delivers it to the TBCD subunit of the Tubulin-binding Complex (TBC). Another tubulin chaperone, TBCB, delivers α-tubulin to the TBCE subunit of the TBC. Previous genetic work in *Aspergillus nidulans* has shown that deletion of *TBCA* causes a near-lethal phenotype, while deletion of *wdA* suppresses this defect, suggesting that TBCA and wdA act in opposite ways to regulate the balance of tubulins. I investigated this balance further by building a construct for overexpression of TBCE, in order to test whether increased TBCE dosage would alter tubulin homeostasis, especially in the context of TBCA loss. To address this question, a TBCE overexpression construct was generated using the inducible *alcA* alcohol dehydrogenase promoter system, which allows controlled expression under different carbon sources. Although the transformation of this construct into *A. nidulans* and the resulting phenotypic analysis will be carried out in future work, the design of the construct establishes the foundation for directly testing the genetic consequences of TBCE overexpression *in vivo*. More broadly, this project is based on the idea that the tubulin cofactor pathway functions as a dynamic homeostatic network in which shifts in the activity of one branch can strongly influence the behavior of the system. The completed construct will allow future experiments to test how increased TBCE expression affects tubulin balance and fungal growth.

**Introduction**

Microtubules organize essential processes in every eukaryotic cell. They support mitosis, intracellular transport, cell polarity, and cell shape. These functions depend on a stable but responsive cytoplasmic pool of assembly-competent αβ-tubulin heterodimers. The cell therefore faces a demanding problem. It must synthesize α-tubulin and β-tubulin, fold them correctly, keep them soluble, assemble them into native heterodimers, and regulate their recycling without allowing harmful imbalances to accumulate. The literature shows that this process does not occur by spontaneous pairing of two freely folded monomers. It proceeds through a dedicated chaperone pathway built around the tubulin binding cofactors TBCA, TBCB, TBCC, TBCD, and TBCE, together with the small GTPase Arl2. This pathway controls the supply, quality, and fate of soluble tubulin, and for that reason sits at the center of microtubule homeostasis (Szolajska and Chroboczek, 2011; Tian and Cowan, 2013; Nithianantham et al., 2015).

This view emerged gradually. Early work had already made clear that tubulin was unusual. Native α-tubulin and β-tubulin were difficult to isolate as stable independent proteins, and bacterial expression produced insoluble material rather than a functional heterodimer. Tian and Cowan brought this evidence together and argued that tubulin biogenesis requires an ordered post-CCT pathway. In their model, newly synthesized tubulin first passes through the cytosolic chaperonin TRiC or CCT forms a double ring chaperonin complex built from eight different but related ATPase subunits. ATP hydrolysis drives repeated conformational changes that support folding of newly synthesized α tubulin and β tubulin. After release from CCT, quasi-native β-tubulin is captured by TBCA or TBCD, and quasi-native α-tubulin is captured by TBCB or TBCE. Exchange between these early and late cofactors produces TBCD-bound β-tubulin and TBCE-bound α-tubulin, and these complexes then converge into a larger assembly intermediate. TBCC promotes the final release step through a GTP-linked reaction involving β-tubulin. This model replaced the old assumption of simple dimer self-assembly with a more precise idea: a cofactor-dependent assembly machine for αβ-tubulin heterodimer formation (Tian and Cowan, 2013).

This pathway was placed in a broader chaperone framework in the study done by Szolajska and Chroboczek (year?). They described tubulin cofactors as highly specific chaperones dedicated to one major client system. Their review emphasized that the cell invests heavily in this pathway because the tubulin heterodimer is abundant, essential, and difficult to produce safely. The pathway includes folding and dimer assembly, dissociation, transient storage, and degradation. This means that tubulin cofactors do not act as simple assembly factors. They regulate tubulin proteostasis across the full life cycle of the heterodimer. This broader view is important because my work examines how changes in one branch of the pathway affect the behavior of the whole system. The genetic interactions described later in this study only make sense if the tubulin cofactor pathway functions as a balanced and reversible network rather than a simple linear assembly route (Szolajska and Chroboczek, 2011).

Within this system, TBCA and TBCB act at the earliest post-chaperonin steps. TBCA binds β-tubulin. TBCB binds α-tubulin. These cofactors stabilize subunit-specific folding intermediates and escort them toward later assembly reactions. In summary, the evidence that TBCA binds free native β tubulin monomers and helps protect cells from β tubulin toxicity (Szolajska and Chroboczek, 2011). Earlier studies in yeast showed that excess β tubulin is more toxic than excess α tubulin because unpaired β tubulin disrupts microtubule organization and interferes with normal cell growth. The TBCA ortholog Rbl2p suppresses this toxicity by <u>binding free β tubulin</u> and limiting its harmful effects. The tubulin-specific chaperone pathway can be driven in reverse: incubation of native heterodimers with a molar excess of TBCD or TBCE results in heterodimer disruption, and when cells are transfected with plasmids engineered for overexpression of TBCD or TBCE, their microtubule network is destroyed (Tian and Cowan, 2013). These findings showed that the cofactor pathway regulates both heterodimer assembly and heterodimer breakdown.

The α-tubulin branch of the pathway centers on TBCB and TBCE. TBCB binds partially folded α-tubulin after release from CCT and delivers it toward TBCE. TBCE acts later in assembly, but TBCE also plays a major role in heterodimer dissociation and α-tubulin turnover. This dual role is central to the focus of this study. Their results established that TBCE is an active dissociation factor and showed that a stoichiometric excess of purified TBCE dissociates tubulin heterodimers *in vitro* (Kortazar et al., 2006). They used TBCA to capture the released β-tubulin monomer, confirming that the disappearance of the tubulin dimer represented true dissociation rather than nonspecific aggregation, and established that the released β-tubulin is recovered as TBCA-β-tubulin complexes. Their results established that TBCE is an active dissociation factor rather than only a forward assembly factor. The study also linked TBCE dysfunction to Kenny Caffey and Sanjad Sakati syndromes, two rare developmental disorders marked by growth defects, skeletal abnormalities, hypoparathyroidism, and impaired neurological development (Kortazar et al., 2006).

Serna et al. (2015) sharpened this picture by examining the complex formed between α-tubulin, TBCE, and TBCB after heterodimer dissociation. Their structural and biochemical work showed that dissociation is efficient only when TBCE and TBCB act together. TBCB alone cannot dissociate the tubulin dimer, and TBCE alone does so only weakly. When both are present, they disrupt the α-tubulin to β-tubulin interface through an energy-independent mechanism requiring no GTP hydrolysis, as confirmed by the finding that the GTP:GDP ratio is unchanged before and after the dissociation reaction. The TBCE CAP-Gly domain binds specifically to the EEY motif at the acidic C-terminal tail of α-tubulin, and this interaction is required for dissociation activity, since removal of the terminal tyrosine residue abolishes the reaction. Cells carrying detyrosinated α tubulin show altered interactions with microtubule associated proteins and defects in normal microtubule behavior, indicating that the EEY motif is important for proper microtubule regulation.

In TBCE the central LRR domain and the linker connecting the UBL domain make additional contacts with α-tubulin that are essential for both dissociation and stabilization of the released α-tubulin monomer, while the UBL domain itself is dispensable for heterodimer disruption. The released α-tubulin is stabilized in a ternary complex with TBCE and TBCB, with the protruding UBL domains in the complex accessible for potential interaction with the proteasome, suggesting a pathway through which the ternary complex may guide α-tubulin toward degradation (Serna et al., 2015). These findings support the idea that the α tubulin branch of the pathway regulates whether tubulin heterodimers remain available for reuse or enter turnover pathways.

A major shift in understanding the late stages of this pathway came from the work of Nithianantham et al. (2015), who reconstituted a stable chaperone complex containing TBCD, TBCE, and Arl2. They showed that these three proteins form a cage-like heterotrimeric complex, termed TBC-DEG, with a hollow core and overall dimensions of approximately 120 × 100 × 90 Å. This complex binds intact soluble αβ-tubulin dimers with high affinity and then recruits TBCC in a nucleotide-dependent manner to form a ternary complex in which Arl2 GTP hydrolysis is activated. Their work shifted the field away from a model in which cofactors interact only transiently. Instead, the late pathway appears organized through stable multi protein complexes that regulate soluble tubulin maintenance and turnover. Critically, a GTP-state-locked Arl2 mutant inhibited ternary complex dissociation *in vitro* and caused severe defects in microtubule dynamics *in vivo*, including increased pausing and loss of rescue events, demonstrating that Arl2 GTPase activity is essential for maintaining proper microtubule dynamics through regulation of the soluble tubulin pool (Nithianantham et al., 2015). This study is important for my thesis because it defines the pathway as a coordinated network with reversible states and shared intermediates, and places TBCE within a larger scaffolded assembly rather than acting as an isolated escort factor.

Taken together, literature supports several conclusions that frame my study. First, tubulin cofactors form a dedicated proteostasis pathway for the soluble αβ-tubulin pool. Second, TBCA and the TBCB-TBCE branch handle different tubulin subunits but remain tightly coupled through shared assembly and dissociation reactions. Third, late cofactors and Arl2 organize larger intermediates that regulate pathway direction and quality control. Fourth, the pathway is inherently reversible, and the net outcome for the cell depends on the balance of activities across all branches simultaneously. These points expose a gap in the current literature. Structural and biochemical studies have identified the pathway components and many of their interactions, but fewer studies have asked how the network behaves genetically in a living fungal cell when one cofactor arm is disrupted, and another is altered through overexpression.

My thesis addresses this gap directly. The key observation motivating this work is that deletion of TBCA in *Aspergillus nidulans* causes a near-lethal phenotype, yet deletion of wdA suppresses this defect and restores growth to near wild-type levels across temperatures where the TBCA deletion strain cannot grow. This genetic interaction suggests that TBCA loss does not create an absolute block in tubulin biogenesis. As TBCA functions in the β tubulin branch of the pathway, suppression by wdA deletion suggested that changes affecting the opposing α tubulin dissociation branch might rebalance the system. Building on this unusual genetic interaction, my thesis examines the consequences of TBCE overexpression on tubulin homeostasis and fungal growth, using a regulated expression system driven by the *alcA* promoter, which is repressed by glucose and induced when ethanol serves as the sole carbon source. Because TBCE occupies a uniquely dual position in the pathway acting in forward α-tubulin delivery during assembly while also driving heterodimer dissociation in partnership with TBCB, its overexpression provides a powerful tool for testing how increased flux through the dissociation branch affects microtubule homeostasis in strains with and without functional TBCA.

My central hypothesis follows this logic. I hypothesize that the severe phenotype caused by loss of TBCA reflects a homeostatic defect in soluble tubulin pool regulation rather than a simple one-step failure in β-tubulin folding. When TBCA is absent, β-tubulin cannot be adequately buffered after release from CCT, and the productive assembly of heterodimers is compromised. Under these conditions, continued or elevated activity of the TBCE-TBCB dissociation branch may worsen the imbalance by releasing α-tubulin from heterodimers faster than new heterodimers can be formed, further depleting the functional tubulin pool. Reducing TBCE activity would therefore suppress this imbalance by slowing dissociation and preserving more soluble heterodimers. by slowing this dissociation flux and allowing more heterodimers to accumulate. I further hypothesize that overexpression of TBCE will exacerbate the phenotype of TBCA deletion by driving excessive heterodimer dissociation, while having a more nuanced or less severe effect in a wild-type background where TBCA can buffer the released β-tubulin. In this model, TBCE overexpression is not simply toxic because it produces too much of one protein. It is toxic in a pathway-specific way, because it shifts the balance of the cofactor network toward dissociation and away from productive assembly.

This is where my work contributes to something new. Earlier studies established the biochemical pathway, resolved stable and transient complexes, and mapped the structural basis of TBCE activity in dissociation. My thesis tests how TBCE overexpression behaves as a genetic perturbation in a filamentous fungal system and asks what this reveals about the balance of the tubulin cofactor network *in vivo*. I will use both deletion and regulated overexpression to infer pathway logic under physiological growth conditions, and I will place TBCE within the pathway not only as an escort factor for α-tubulin but as a regulator of reversible flux whose dosage becomes especially important when β-tubulin buffering by TBCA is lost.

I argue that the tubulin cofactor pathway in *Aspergillus nidulans* functions as a balanced and reversible homeostatic network rather than a one direction folding pathway. Loss of TBCA disrupts β tubulin balance and causes severe growth defects. Altering the activity of the TBCE branch changes this balance by affecting the rate of heterodimer dissociation. By testing TBCE overexpression in vivo, my work examines how changes in cofactor dosage influence the stability of the soluble tubulin pool and microtubule homeostasis.

## Materials and Methods

### Study design

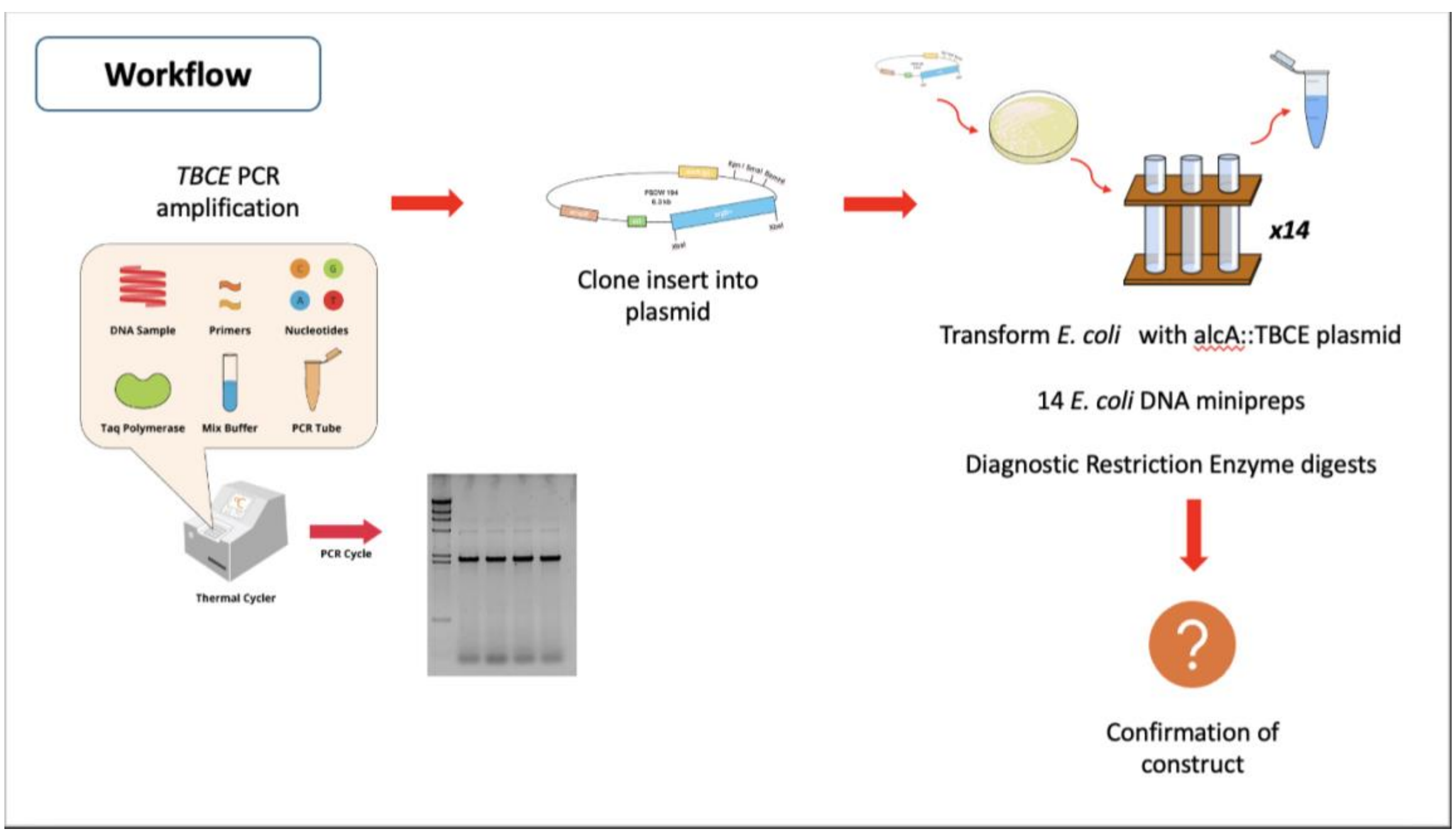

Figure 1: The overall workflow of the project.

An ethanol-inducible TBCE overexpression system was constructed for later transformation into *Aspergillus nidulans*. The cloning workflow included primer design, PCR amplification of the TBCE genomic fragment, restriction digestion of insert and vector, dephosphorylation of the vector backbone, ligation, transformation into *E. coli* - DH5α competent cells, plasmid miniprep from ampicillin-resistant colonies, and diagnostic restriction digest analysis by agarose gel electrophoresis. The vector used in this work was pSDW194, an *E. coli* to *A. nidulans* shuttle plasmid carrying the *alcA* alcohol dehydrogenase promoter, argB, an origin of replication, and ampicillin resistance. The final construct placed TBCE downstream of the *alcA* promoter for inducible expression.

**Primer design**

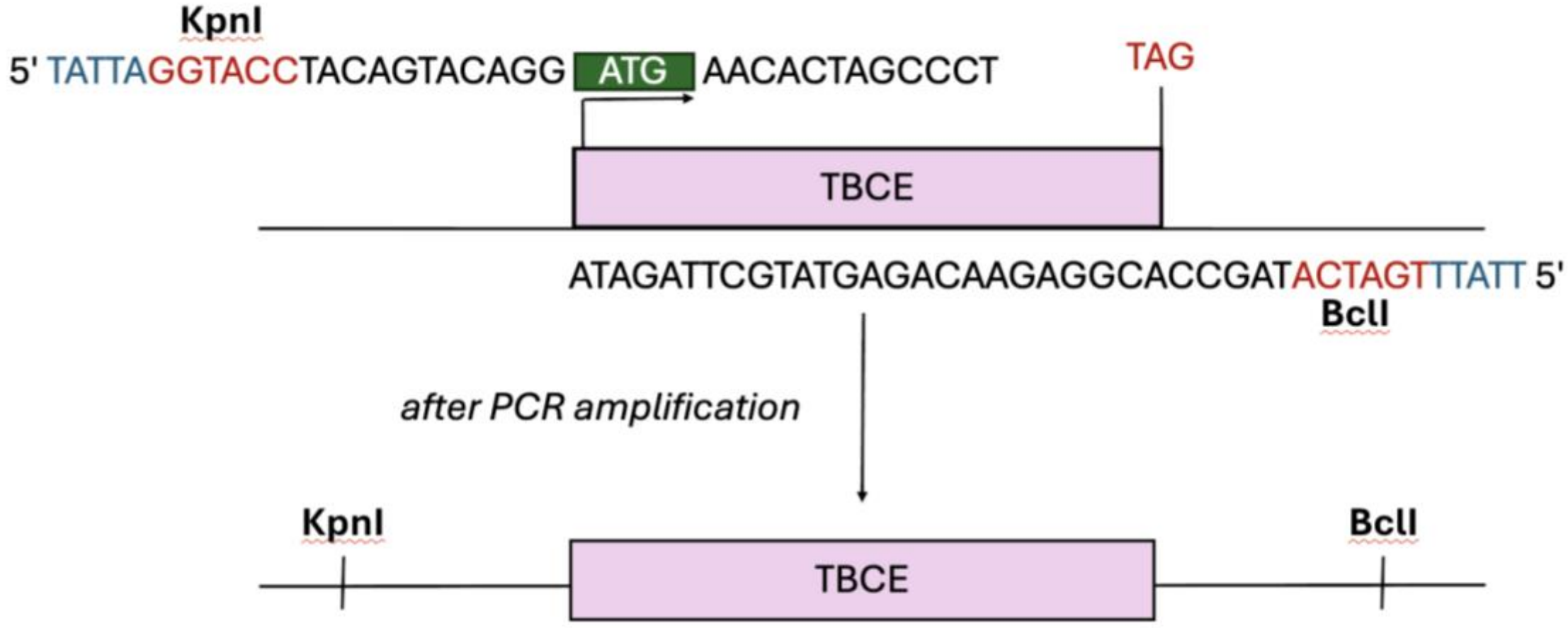

Figure 2: The strategy used to amplify cloning TBCE. For PCR amplification, primers were designed to amplify the TBCE gene and add restriction sites to the ends. The forward primer added a KpnI site, and the reverse primer added a BclI site. These restriction sites are important because they allow the amplified TBCE product to be cut and inserted into the plasmid in a controlled way.

Primers were designed from the *Aspergillus nidulans* TBCE genomic sequence using the same workflow used in the molecular genetics primer design protocol. The TBCE genomic sequence was obtained from FungiDB and analyzed in SerialCloner 2.1 and IDT OligoAnalyzer for melting temperature (Tm), hairpins, self-dimers, cross dimers, palindromes, and nucleotide run. A KpnI site was built into the forward primer and a BclI site into the reverse primer to support directional cloning into pSDW194. The forward primer sequence was 5′ TATTAGGTACCATCTACAGTACAGGATGAACACTAGCCCT 3′. The reverse primer sequence was 5′ TTATTTGATCATAGCCCACGGAGAACAGAGTATGCTAGATA 3′. The forward primer annealed near nucleotides 2987 to 3015 of the TBCE region. The reverse primer annealed near nucleotides 5116 to 5087. This primer pair was expected to produce a 2129 bp amplicon. The recorded melting temperatures were 70.3°C for the forward primer and 69.9°C for the reverse primer. The forward primer included five nucleotides upstream of the KpnI site, followed by 5′ UTR sequence, the ATG start codon, and downstream coding sequence. The reverse primer included five nucleotides upstream of the BclI site and extended to 115 nt downstream of the TBCE stop codon.

Table 1. TBCE primer sequences and amplification parameters used for cloning.

| Primer | Sequence (5′ → 3′) | Restriction Site | Tm (°C) | Primer Length | Expected Amplicon |
|---|---|---|---|---|---|
| alcA-TBCE KpnI Forward | TATTAGGTACCATCTACAGTACAGGATGAACACTAGCCCT | KpnI | 70.3 | 40-mer | 2129 bp |
| alcA-TBCE BclI Reverse | TTATTTGATCATAGCCCACGGAGAACAGAGTATGCTAGATA | BclI | 69.9 | 41-mer | 2129 bp |

**PCR amplification of TBCE**

TBCE was amplified from *Aspergillus nidulans* genomic DNA from strain PCS 439, genotype riboA1 yA2. PCR amplification was performed using Q5 High Fidelity DNA Polymerase to minimize sequence errors during amplification. Genomic DNA from Aspergillus nidulans served as template DNA. PCR reactions were assembled on ice in sterile PCR tubes. Each 50 µL PCR reaction contains 10 µL of 5X Q5 Reaction Buffer, 1 µL of 10 mM dNTP mix, 2.5 µL of 10 µM forward primer, 2.5 µL of 10 µM reverse primer, 1 µL genomic DNA template, 0.5 µL Q5 High Fidelity DNA Polymerase, and 32.5 µL nuclease free water. PCR reactions were amplified using the following thermal cycler conditions: initial denaturation at 98°C for 30 seconds, followed by 30 cycles of denaturation at 98°C for 10 seconds, primer annealing at 65°C for 10 seconds and extension at 72°C for approximately 2 minutes. A final extension step was carried out at 72°C for 2 minutes before the reaction was held at 4°C.

**Agarose gel analysis of PCR products**

The PCR reactions were checked by agarose gel electrophoresis. A 1% agarose gel was prepared in 1x TAE buffer. λ HindIII DNA was loaded as the size marker. The relevant marker bands were 23.1 kb, 9.4 kb, 6.6 kb, 4.4 kb, 2.3 kb, 2.0 kb, and 0.56 kb. PCR samples were loaded into adjacent lanes. The gel was run until the dye front migrated far enough to resolve the product near the 2.0 to 2.3 kb region. The gel was stained with Gel Red and imaged on the ChemiDoc system.

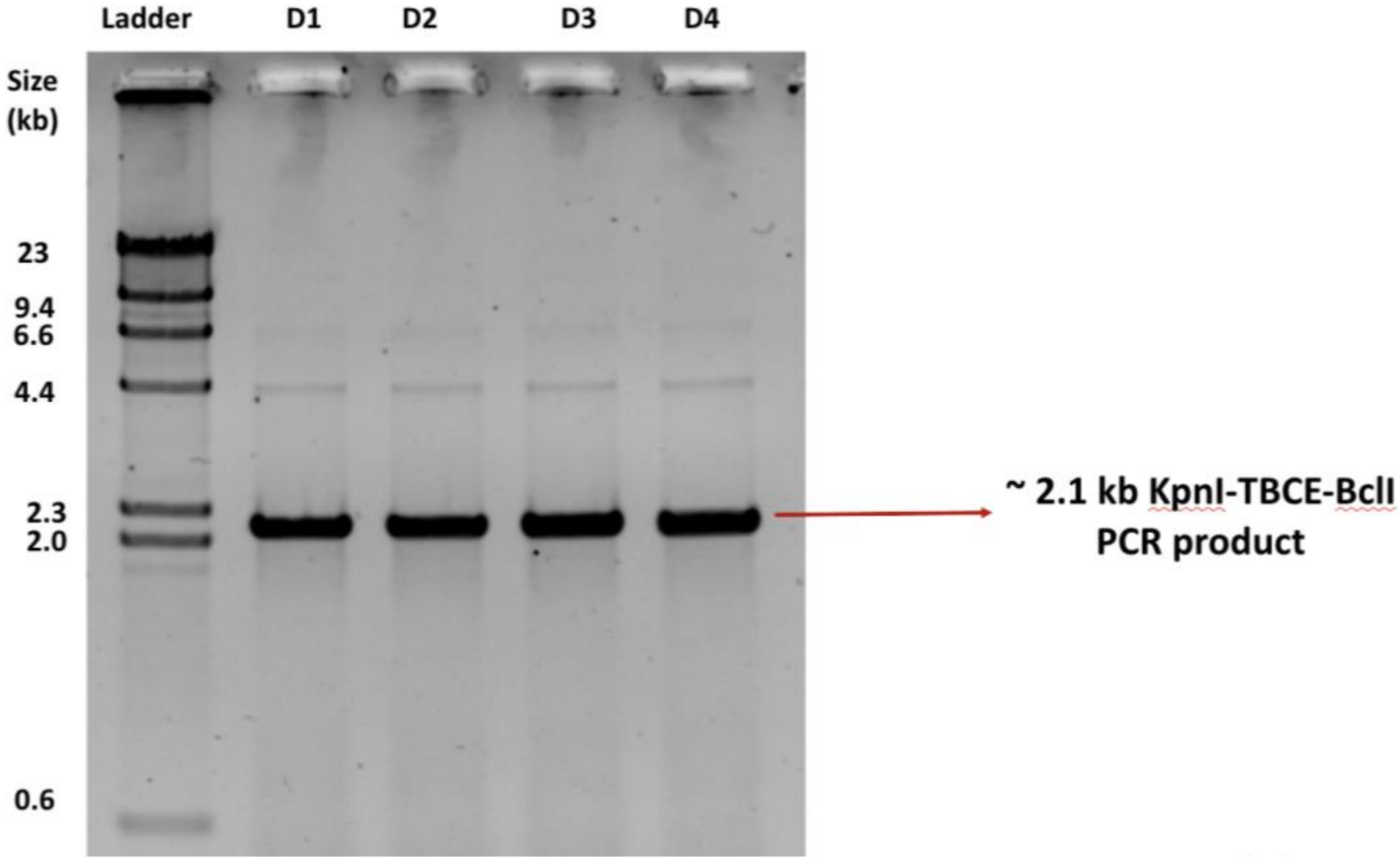

Figure 3: Agarose gel showing PCR amplification of TBCE. The arrow marks the expected product at approximately 2.1 kb.

**Gel extraction and purification of PCR products**

PCR amplicons corresponding to the expected 2129 bp TBCE product were excised from the agarose gel using a sterile razor blade under blue light illumination. Gel slices were transferred to pre-weighed microcentrifuge tubes and weighed to determine gel mass. DNA was purified from excised gel fragments using the GeneJET Gel Extraction Kit according to the laboratory protocol. Binding buffer was added to each gel slice according to gel mass and samples were incubated at 50 to 55°C until the agarose dissolved completely. Dissolved samples were transferred to silica spin columns and centrifuged to bind DNA to the column matrix. Columns were washed with the supplied wash buffer containing ethanol to remove salts, residual agarose, nucleotides, primers, and Q5 PCR reaction buffer components that would interfere with downstream restriction digestion reactions.

Removal of the residual Q5 PCR buffer was important because salts and buffer components remaining from the PCR reaction reduce restriction enzyme efficiency. The PCR amplicon was purified using the GeneJET silica column purification system before restriction digestion. During purification, the PCR mixture was passed through the silica membrane, washed repeatedly with the ethanol containing wash buffer, and eluted in nuclease free water. This purification step removed the original Q5 reaction buffer so that the purified amplicon could be digested efficiently in the restriction enzyme buffer used for KpnI and BclI digestion. After washing, columns were centrifuged an additional time to remove residual ethanol. DNA was eluted in nuclease free water and quantified using a NanoDrop spectrophotometer.

**Restriction digestion of vector and insert DNA**

The purified TBCE PCR amplicon and the pSDW194 shuttle plasmid were digested with KpnI and BclI restriction enzymes to generate compatible cohesive ends for directional cloning downstream of the *alcA* promoter.

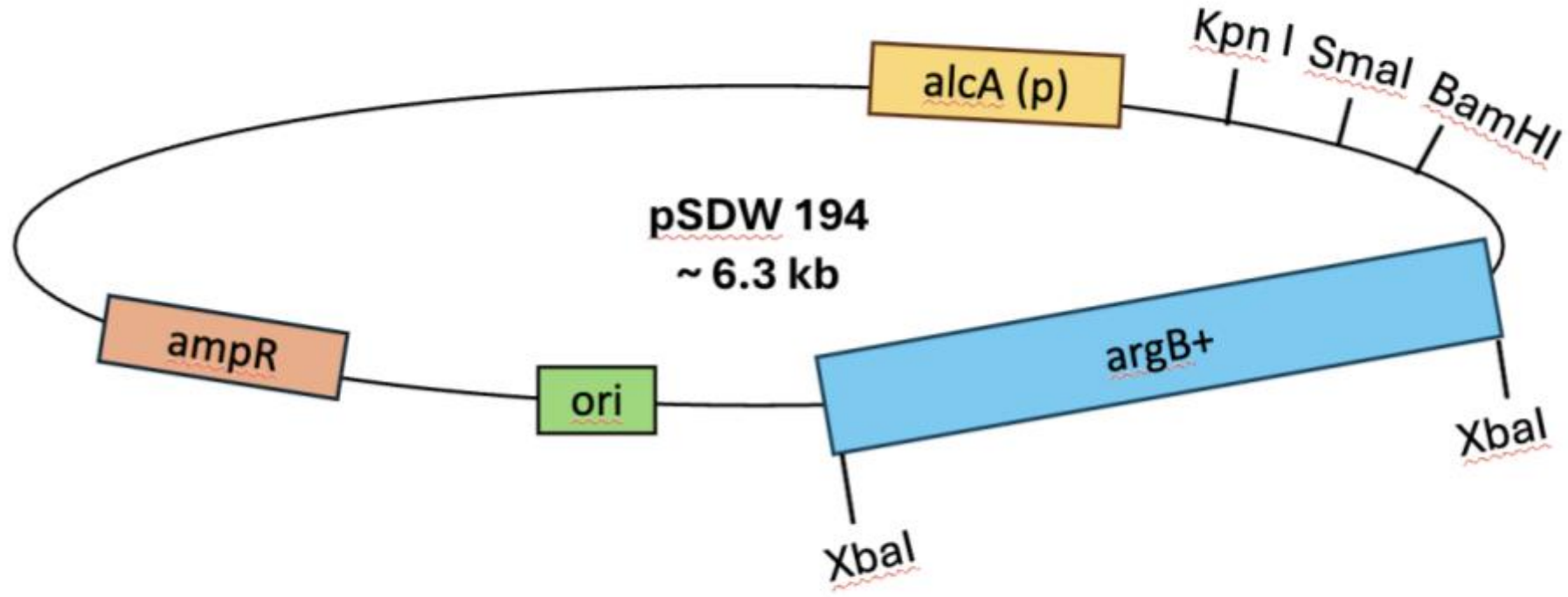

Figure 4: Plasmid map.

The forward primer introduced a KpnI recognition site at the 5′ end of the amplicon and the reverse primer introduced a BclI recognition site at the 3′ end. Digestion with KpnI produced a cohesive 3′ GTAC overhang at the 5′ side of the TBCE amplicon. Digestion with BclI produced a cohesive 5′ GATC overhang at the opposite end of the amplicon. Digestion of the plasmid vector with the same enzymes generated complementary overhangs and forced the insert into a single orientation during ligation downstream of the *alcA* promoter.

Restriction digest reactions were assembled using purified DNA, 10X NEB CutSmart restriction buffer, KpnI, BclI, and nuclease free water in a final volume of 50 μL. Digest was incubated at 37°C for approximately 2 hours. After digestion, samples were loaded onto a 1% agarose gel prepared in 1X TAE buffer. The linearized pSDW194 vector band and the approximately 2.1 kb TBCE insert band were excised carefully using a sterile razor blade. DNA fragments were purified again using the GeneJET Gel Extraction Kit in order to remove restriction enzymes, salts, and small DNA fragments generated during digestion. The pSDW194 plasmid used in this study was approximately 6.3 kb in size and contained the inducible *alcA* promoter, the argB selectable marker for fungal transformation, the ampicillin resistance cassette for bacterial selection, and a bacterial origin of replication. The cloning region downstream of the *alcA* promoter contained KpnI and BamHI restriction sites used for insertion of the TBCE fragment.

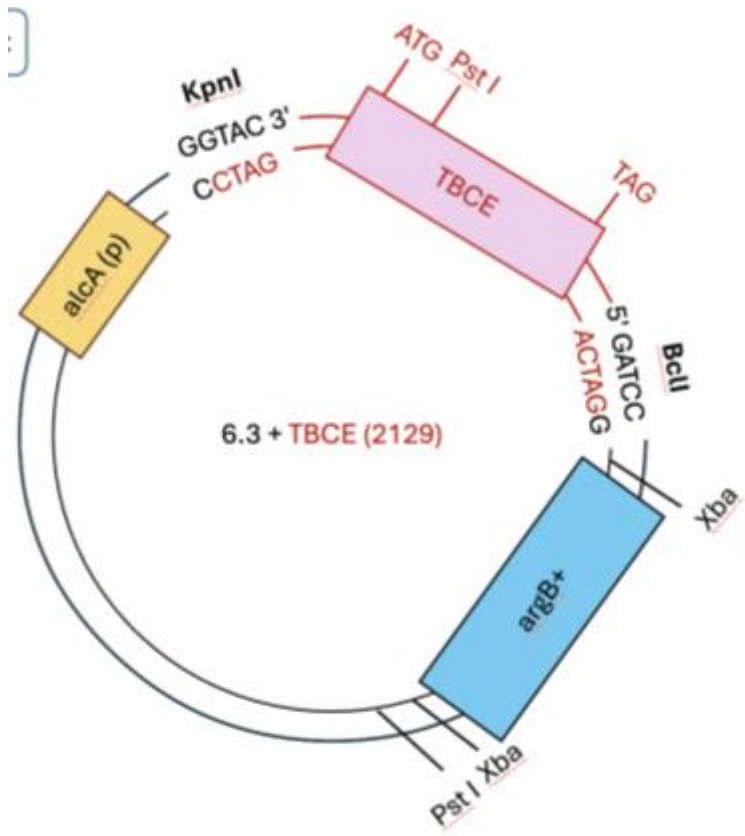

Figure 5: Cloning strategy for construction of the alcA::TBCE overexpression plasmid. The TBCE PCR fragment containing engineered KpnI and BclI restriction sites was inserted downstream of the *alcA* promoter in the pSDW194 shuttle plasmid. Directional cloning was achieved using compatible cohesive ends generated by KpnI, BclI, and BamHI digestion.

**DNA ligation**

Purified digested vector and insert DNA were ligated using T4 DNA Ligase. Vector and insert concentrations were first measured using NanoDrop spectrophotometry, and pmole amounts were calculated using fragment size and DNA concentration. The ligation reaction was assembled using approximately a 1:3 molar ratio of vector to insert DNA. Approximately 0.02 pmoles of digested pSDW194 vector DNA and 0.06 pmoles of digested TBCE insert DNA were combined in the ligation mixture. The final ligation volume was 20 µL. Each reaction contained digested vector DNA, digested insert DNA, 2 µL of 10X T4 DNA ligase buffer containing ATP, 1 µL T4 DNA ligase, and nuclease free water to final volume. Ligation reactions were incubated overnight at 16°C.

**Transformation into *E. coli* DH5α**

Chemically competent *E. coli* DH5α cells were transformed with the ligation product using heat shock transformation. Aliquots of competent cells were thawed on ice and mixed gently with ligation reaction DNA. Cells were incubated on ice for 30 minutes; heat shocked at 42°C for 45 seconds and immediately returned to ice for 2 minutes. SOC recovery medium was added, and cells were incubated at 37°C with shaking for 1 hour before plating. Transformed cells were plated onto LB agar plates containing ampicillin and incubated overnight at 37°C.

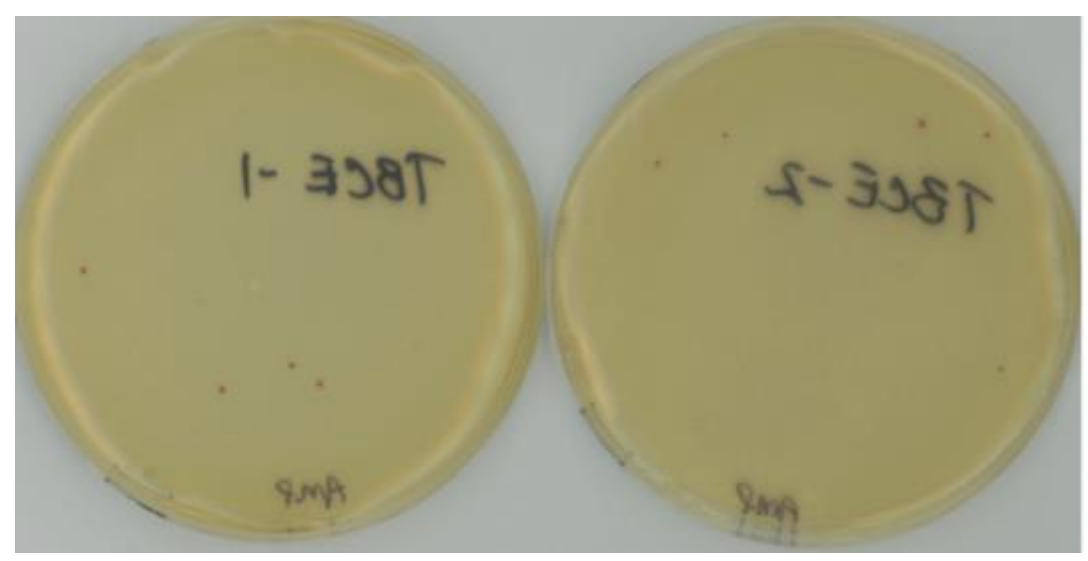

Figure 6: Ampicillin-resistant *E. coli* transformants carrying candidate alcA::TBCE plasmids.

## Plasmids miniprep

Individual ampicillin resistant colonies were inoculated into 5 mL LB or 2xYT medium containing 100 µg/mL ampicillin and incubated overnight at 37°C with vigorous shaking. Approximately 3 mL of overnight culture was collected by centrifugation, and bacterial pellets were resuspended in 110 µL ice cold GTE buffer containing 50 mM glucose, 10 mM EDTA, and 25 mM Tris HCl, pH 8.0. Cells were incubated briefly at room temperature before alkaline lysis with freshly prepared NaOH and SDS solution. Tubes were mixed gently by inversion and incubated on ice. Neutralization was performed using ice cold potassium acetate solution and cellular debris was pelleted by centrifugation. Supernatants were transferred carefully into fresh tubes to avoid carryover of precipitated debris. Plasmid DNA was extracted using

phenol:chloroform:isoamyl alcohol followed by chloroform extraction. DNA was precipitated using sodium acetate and cold ethanol, washed twice with 70% ethanol, dried, and resuspended in TE buffer containing RNase A. During the last preparation step, excess RNase A (750ul) was added accidentally to a sample. The DNA pellet was therefore dried again, and the purification steps were repeated before final resuspension. Purified plasmid DNA samples were quantified using NanoDrop spectrophotometry and used for diagnostic restriction digest analysis.

**Diagnostic restriction digest analysis**

Candidate plasmids were screened by diagnostic restriction digestion to confirm successful insertion of the TBCE fragment into the pSDW194 vector. Restriction mapping predicted that digestion of the completed alcA::TBCE plasmid with PstI would generate two fragments of approximately 4.97 kb and 3.46 kb. The expected fragment sizes were determined from the plasmid map and restriction digest strategy prepared before plasmid screening. Individual plasmid minipreps were digested with PstI in reactions containing plasmid DNA, 10X restriction buffer, PstI enzyme, and nuclease free water. Reactions were incubated at 37°C and then analyzed on agarose gels alongside λ HindIII and pSWD194 molecular weight standards. Fourteen plasmid candidates were originally prepared for screening. Thirteen samples were loaded onto the final diagnostic gel because there was insufficient space on the gel to include all fourteen lanes together with the ladder controls. Reactions were incubated at 37°C and then analyzed on agarose gels alongside λ HindIII and pSWD194 molecular weight standards.

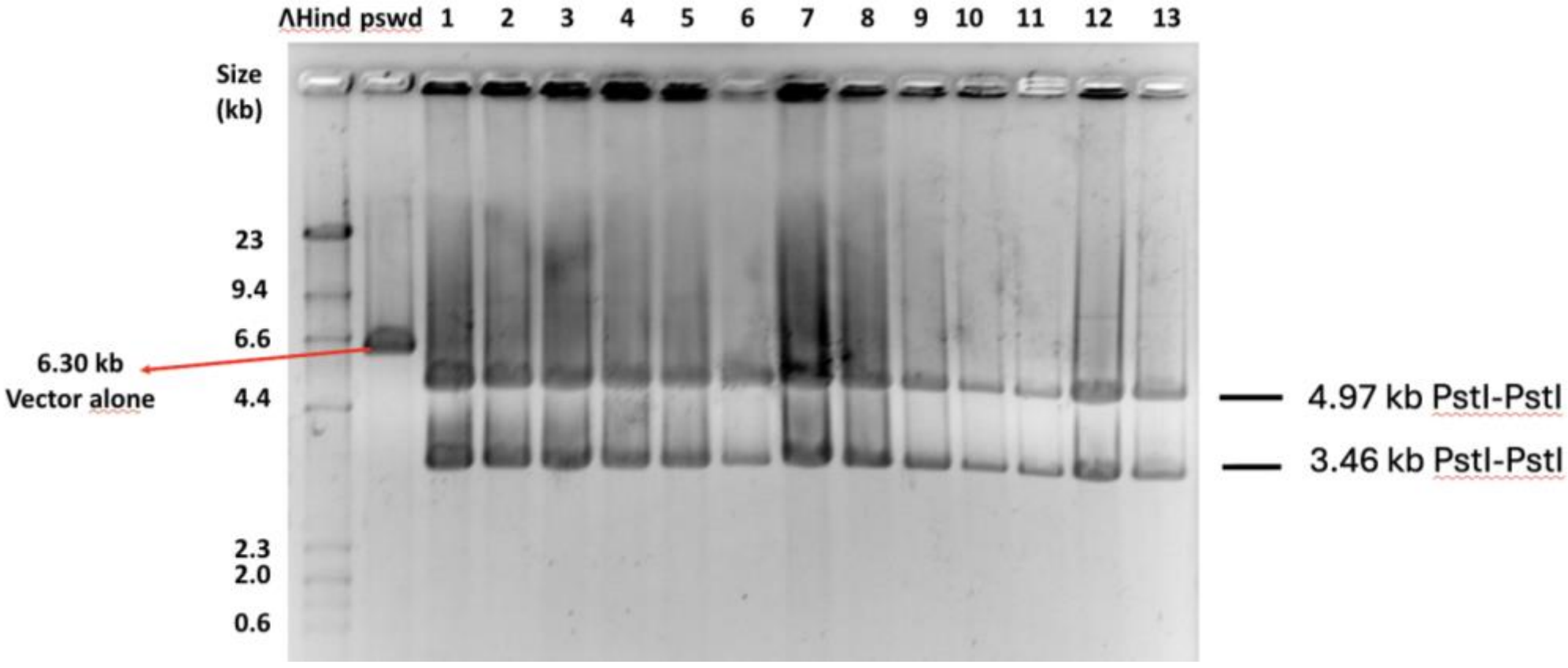

Figure 7: Diagnostic restriction digest of candidate plasmids. Band sizes are consistent with the predicted alcA::tbcE construct.

## Results

PCR amplification of the TBCE genomic fragment produced a band near the expected 2.1 kb size on agarose gels. The amplified fragment matched the predicted 2129 bp KpnI-TBCE-BclI product produced from the primer design strategy. λ HindIII DNA was used as the molecular weight marker. The amplified bands appeared concentrated and finalized relative to the 2.0 kb and 2.3 kb marker fragments. The sample lanes showed very little non-specific amplification. These results demonstrated that the primer pair amplified the targeted TBCE region from the genomic DNA of *Aspergillus nidulans* and made enough DNA for further cloning processes. The PCR amplification products and anticipated fragment size are displayed in Figure 3, which is 2.1kb. After PCR amplification, the TBCE amplicon was excised from the agarose gel and purified using the GeneJET silica column purification system. This purification step removed agarose, salts, primers, residual nucleotides, and Q5 reaction buffer components before restriction digestion. The purified TBCE insert and pSDW194 vector were then digested with the appropriate restriction enzymes and prepared for ligation into the *alcA* expression vector.

The purified TBCE insert and pSDW194 plasmid backbone were ligated using T4 DNA ligase to generate the alcA::TBCE construct. The cloning strategy placed TBCE downstream of the ethanol inducible *alcA* promoter in the pSDW194 shuttle vector. KpnI digestion generated a GTAC cohesive overhang at the 5′ end of the insert. BclI digestion generated a GATC cohesive overhang at the opposite end of the insert. Digestion of the vector with KpnI and BamHI generated compatible cohesive ends for directional cloning. This strategy forced the TBCE fragment into a single orientation downstream of the *alcA* promoter. The expected recombinant plasmid size was approximately 8.4 kb, consisting of the 6.3 kb pSDW194 backbone and the 2129 bp TBCE insert.

Transformation of the ligation products into chemically competent *E. coli* DH5α cells produced multiple ampicillin resistant colonies on 2xYT agar plates containing 100 µg/mL ampicillin. Small, isolated colonies were selected for plasmid screening. 14 colonies were inoculated into liquid culture and processed for plasmid miniprep. Plasmid DNA isolated from these cultures produced detectable DNA concentrations by NanoDrop spectrophotometry and was carried forward for diagnostic restriction digest analysis. Diagnostic screening of candidate plasmids was performed using PstI restriction digestion. Restriction mapping predicted that successful alcA::TBCE constructs would generate two DNA fragments of approximately 4.97 kb and 3.46 kb following digestion. 13 plasmid samples were loaded onto the final agarose gel together with λ HindIII molecular weight markers and vector controls. One candidate plasmid was skipped because the gel did not contain sufficient space to fit all 14 samples together with the controls. All 13 analyzed plasmids produced restriction fragment patterns consistent with the predicted alcA::TBCE construct. Each sample generated bands near the expected 4.97 kb and 3.46 kb fragment sizes. The vector only control produced a distinct banding pattern near the 6.3 kb plasmid size and lacked the predicted insert associated fragments.

## Discussion

The primary goal of this experiment was to construct an inducible TBCE overexpression system for later transformation into *Aspergillus nidulans*. I successfully amplified the TBCE genomic region, cloned the fragment into the pSDW194 vector downstream of the *alcA* promoter, and recovered candidate plasmids with restriction digest patterns consistent with the predicted alcA::TBCE construct. The completed plasmid will be used in future *Aspergillus nidulans* transformation experiments to examine how increased TBCE expression affects microtubule homeostasis.

PCR amplification was the first major step in constructing the overexpression system. The primer pair produced a strong DNA band near the predicted 2129 bp size. The primers were designed to introduce restriction sites for directional cloning while retaining the coding region of TBCE. The forward primer had a KpnI site, and the reverse primer had a BclI site. This design allowed the insert to ligate directionally into the plasmid downstream of the *alcA* promoter. The agarose gel showed a distinct band near the expected size with little nonspecific amplification, which made the PCR product suitable for cloning.

The cloning strategy used compatible restriction enzyme overhangs to insert the TBCE fragment into the plasmid in the correct orientation. Digestion with KpnI created a GTAC overhang at one end of the insert, while digestion with BclI created a GATC overhang that matched the BamHI digested site in the pSDW194 vector. This allowed the TBCE fragment to ligate downstream of the *alcA* promoter. The *alcA* promoter system was selected because it provides inducible expression in *Aspergillus nidulans* under ethanol inducing conditions. That means when cells are grown under inducing conditions, TBCE expression should increase. When cells are grown under non-inducing conditions, TBCE expression should remain lower. This system will allow future experiments to test how increased TBCE expression affects fungal growth, microtubule organization, and interactions with wdA

The diagnostic PstI digest was the final step used to check plasmid construction. Restriction mapping predicted two fragments of approximately 4.97 kb and 3.46 kb if the TBCE insert ligated correctly into the plasmid backbone. All thirteen screened plasmids produced digestion patterns consistent with these predicted fragment sizes, which means they are competent cells. The vector control produced a different banding pattern near the original 6.3 kb plasmid size and lacked the insert fragment.

This construct creates an experimental tool for future analysis of tubulin cofactor mediated microtubule regulation. Previous work in the James laboratory showed that loss of wdA causes severe cytoskeletal defects including absence of microtubules, mitotic catastrophe, and cold sensitive lethality. Additional observations showed that deletion of TBCA produced a severe growth defect that was rescued in strains lacking wdA. These genetic interactions suggest that wdA is connected to pathways involved in tubulin cofactor regulation and microtubule homeostasis. Since TBCE functions in α-tubulin folding and heterodimer assembly, changing TBCE expression levels may affect the balance between soluble tubulin and microtubule stability.

To conclude, this project successfully assembled an inducible alcA::TBCE overexpression construct. PCR amplification and diagnostic restriction analysis supported correct plasmid construction. This construct provides a tool for future transformation into *Aspergillus nidulans* to investigate how TBCE overexpression affects cell growth and microtubule function. Future comparisons between wild-type and wdA mutant strains under inducing conditions may reveal genetic interactions between wdA and the tubulin cofactor pathway. Overall, this work contributes to understanding how cells regulate tubulin folding, availability, and microtubule stability.